\documentclass[10pt,prd,aps,twocolumn,showpacs,nofootinbib,superscriptaddress,floatfix,nopreprintnumbers]{revtex4}

\usepackage{graphicx}
\usepackage{bm} 
\usepackage{subfigure}

\usepackage[hidelinks]{hyperref}

\begin{document}
 
\title{Anisotropic hydrodynamics for mixture of quark and gluon fluids}

\author{Wojciech Florkowski} 
\affiliation{The H. Niewodnicza\'nski Institute of Nuclear Physics, Polish Academy of Sciences, PL-31342 Krak\'ow, Poland}

\author{Ewa Maksymiuk} 
\affiliation{Institute of Physics, Jan Kochanowski University, PL-25406~Kielce, Poland} 

\author{Radoslaw Ryblewski} 
\affiliation{The H. Niewodnicza\'nski Institute of Nuclear Physics, Polish Academy of Sciences, PL-31342 Krak\'ow, Poland} 

\author{Leonardo Tinti} 
\affiliation{Institute of Physics, Jan Kochanowski University, PL-25406~Kielce, Poland}

\date{\today}

\begin{abstract}
A system of equations for anisotropic hydrodynamics is derived that describes a mixture of anisotropic quark and gluon fluids. The consistent treatment of the zeroth, first and second moments of the kinetic equations allows us to construct a new framework with more general forms of the anisotropic phase-space distribution functions than those used before. In this way, the main difficiencies of the previous formulations of anisotropic hydrodynamics for mixtures have been overcome and the good agreement with the exact kinetic-theory results is obtained.
\end{abstract}

\pacs{25.75.-q, 25.75.Dw, 25.75.Ld}

\keywords{relativistic heavy-ion collisions, quark-gluon plasma, relativistic hydrodynamics}

\maketitle 

\section{Introduction}
\label{sect:intro}

The successful description of relativistic heavy-ion collisions at RHIC and at the LHC in terms of relativistic dissipative fluid dynamics (for a recent review see \cite{Heinz:2013th}) has brought a lot of attention to studies aiming at the construction of the most adequate hydrodynamic framework. One way to achieve this task is to compare the results of various hydrodynamic approaches \cite{Koide:2006ef, Muronga:2006zx, Baier:2007ix, Bhattacharyya:2008jc, Betz:2008me, PeraltaRamos:2009kg, El:2009vj, 
Denicol:2010xn, Martinez:2010sc, Florkowski:2010cf,Strickland:2014pga}, which differ by the number of terms included in the formalism and by the values of the transport coefficients, with the results of the underlying microscopic kinetic theory  \cite{Florkowski:2013lza,Florkowski:2013lya,Florkowski:2014sfa,Florkowski:2014sda,Denicol:2014xca,Denicol:2014tha}. The latter is very often used as a staring point to derive the specific form of the evolution equations of relativistic hydrodynamics, however, several approximations done in such procedures may result in differences between the predictions of the kinetic theory and the hydrodynamic models constructed directly with its help.

As a lot of work has been already done in this context for simple (i.e., one-component) fluids, the analysis of mixtures has been so far quite limited \cite{Florkowski:2012as,Florkowski:2013uqa,Florkowski:2014txa}, for some recent developments see~\cite{Jaiswal:2015mxa,Kikuchi:2015swa,Rougemont:2015ona}.  One of the problems of the previous approaches using the concept of anisotropic hydrodynamics~\cite{Florkowski:2012as,Florkowski:2013uqa,Florkowski:2014txa} was that they were based only on the zeroth and first moments of the kinetic equations. Assuming that the distribution functions used in anisotropic hydrodynamics are described by the original Romatschke-Strickland form~\cite{Romatschke:2003ms}, one finds underdetermined set of equations,  where the number of unknown parameters is larger than the number of equations. Consequently, to close the system of equations, in Refs.~\cite{Florkowski:2012as,Florkowski:2013uqa,Florkowski:2014txa}  the transverse-momentum scale parameters for quarks and gluons were taken to be equal~\footnote{The transverse-momentum scale parameters can be interpreted also as transverse temperatures --- parameters characterising  transverse-momentum distributions. }. 

In this work we develop the approach presented in Refs.~\cite{Florkowski:2012as,Florkowski:2013uqa,Florkowski:2014txa}. We use  the zeroth, first, as well as the second moments of the kinetic equations for quarks, antiquarks and gluons. This allows us to use the Romatschke-Strickland form with more independent parameters as compared to the previous works. Our selection of the equations will be presented and discussed in greater detail below. Here we only mention that our equations include the baryon number and energy-momentum conservation laws. This is achieved by the use of the two types of the Landau matching condition: the first one fixes the effective chemical potential $\mu$, while the second one (more commonly used) fixes the effective temperature $T$. Moreover, we use a special combination of equations coming from the second moment which guarantees the agreement with the Israel-Stewart theory for the system approaching local equilibrium~\cite{Tinti:2013vba}, see also~\cite{Florkowski:2014bba,Nopoush:2014pfa,Nopoush:2015yga}.  

Our approach is restricted to the one-dimensional, boost invariant systems~\cite{Bjorken:1982qr},  denoted below as (0+1)D systems.  We show that the evolution equations of anisotropic hydrodynamics for quark and gluon fluids derived herein yield good agreement with the results  of the kinetic theory. Several options for the selection of the zeroth moment equations have been studied and the best choice is indicated.  The new approach eliminates solutions with exponentially damped anisotropy parameters, found in~\cite{Florkowski:2012as}, which do not agree with the kinetic-theory solutions~\cite{Florkowski:2014txa}. Such solutions appeared in the cases where the initial conditions corresponded to oblate-prolate or prolate-prolate initial quark and gluon momentum distribution functions. The new solutions have the same qualitative character for all types of the initial conditions (oblate-oblate, oblate-prolate, and prolate-prolate), do not exhibit exponential damping, and closely follow the kinetic-theory solutions for systems far from and close to local thermal equilibrium.

The paper is written as follows: In Sec.~\ref{sect:ke} the kinetic equations for quarks, antiquarks, and gluons in the relaxation time approximation are introduced. In Sec.~\ref{sect:zeromoment} the zeroth moments of the quark and antiquark equations are discussed in the context of the baryon number conservation. In Sec.~ \ref{sect:firstmoment} we discuss the first moment of the kinetic equations and analyse the energy-momentum conservation. The second moments are discussed in Sec.~\ref{sect:secondmoment} and our results are presented in Sec.~\ref{sect:res}. We conclude in Sec.~\ref{sect:con}. Throughout the paper we use natural units and the metric tensor's signature is $(+,-,-,-)$.

\section{Kinetic equations}
\label{sect:ke}

We start our analysis with the kinetic equations for quarks, antiquarks and gluons written in the relaxation time approximation (RTA)~\cite{Bhatnagar:1954zz,aw,book}. They read
\begin{equation}
 p^{\mu }\partial_{\mu } Q^\pm (x,p)= 
- p^\mu U_\mu \frac{Q^\pm(x,p) - Q^\pm_{\rm eq}(x,p)}{\tau_{\rm eq}},  
\label{kineq0}
\end{equation}
\begin{equation}
p^{\mu }\partial_{\mu } G(x,p) = 
- p^\mu U_\mu \frac{G(x,p) - G_{\rm eq}(x,p)}{\tau_{\rm eq}},
\label{kineg0}
\end{equation} 
where $Q^+(x,p)$ ($Q^-(x,p)$) is the quark (antiquark) phase-space distribution function, $G(x,p)$ is the gluon distribution function, and $\tau_{\rm eq}$ is the relaxation time. The four-vector $U$ describes the hydrodynamic flow in the Landau frame (i.e., $U$ is defined as the eigen four-vector of the energy-momentum tensor).

The quark and gluon distribution functions are assumed to have a generic structure~\cite{Romatschke:2003ms} 
\begin{eqnarray}
Q^\pm(x,p) &=& \exp\left(\frac{\pm \lambda_q - \sqrt{(p\cdot U)^2 + \xi_q (p\cdot Z)^2}}{\Lambda_q} \right),
\nonumber \\
G(x,p) &=& \exp\left(-\frac{\sqrt{(p\cdot U)^2 + \xi_g (p\cdot Z)^2}}{\Lambda_g} \right),
\label{RSform}
\end{eqnarray}
where the parameters $\Lambda_q$ and $\Lambda_g$ define the transverse momentum scale, $\lambda_q$ is the non-equilibrium baryon chemical potential of quarks, while $\xi_q$ and $\xi_g$ are the anisotropy parameters. In local equilibrium, the two anisotropy parameters vanish and Eqs.~(\ref{RSform}) are reduced to the standard equilibrium distributions
\begin{eqnarray}
Q^\pm_{\rm eq}(x,p) &=& \exp\left(\frac{\pm \mu - p\cdot U}{T}  \right),
\nonumber \\
G_{\rm eq}(x,p) &=& \exp\left(-\frac{p\cdot U}{T}  \right),
\label{eqforms}
\end{eqnarray}
where $T$ is the temperature and $\mu$ is the baryon chemical potential. For the sake of simplicity, we assume here the classical Boltzmann statistics. A generalisation of the present results to the case of the quantum Bose-Einstein and Fermi-Dirac statistics is straightforward~\cite{Florkowski:2014sda,Florkowski:2015lra}. 

The equilibrium distribution functions of the form  (\ref{eqforms}) are used to define the RTA collision terms in~(\ref{kineq0}) and (\ref{kineg0}). In this case $\mu$ and $T$ should be treated as the {\it effective} baryon chemical potential and effective temperature that are determined by the appropriate Landau matching conditions.

In addition to the flow vector $U$, that can be parameterised in terms of the three-velocity in the standard way as
\begin{equation}
U^\mu = \gamma (1, v_x, v_y, v_z), \quad \gamma=(1-v^2)^{-1/2},
\label{U}
\end{equation}
we introduce the four-vector $Z$ defined as~\cite{Florkowski:2010cf}
\begin{equation}
Z^\mu = \gamma_z (v_z, 0, 0, 1), \quad \gamma_z = (1-v_z^2)^{-1/2}.
\label{Z}
\end{equation}
The appearance of $Z$ is connected with the privileged direction of the beam axis. 

The four vectors $U$ and $Z$ satisfy the following normalisation conditions
\begin{eqnarray}
U^2 = 1, \quad Z^2 = -1, \quad U \cdot Z = 0.
\label{UZnorm}
\end{eqnarray}
In the local rest frame of the fluid element, $U^\mu$ and $Z^\mu$ have simple forms
\begin{eqnarray}
 U^\mu = (1,0,0,0), \quad Z^\mu = (0,0,0,1). 
 \label{UZLRF}
\end{eqnarray}
In the (0+1)D case, we may further use
\begin{eqnarray}
 U^\mu &=& (t/\tau,0,0,z/\tau), \nonumber \\
 Z^\mu &=& (z/\tau,0,0,t/\tau),
 \label{UZbinv}
\end{eqnarray}
where $\tau$ is the (longitudinal) proper time
\begin{eqnarray}
\tau = \sqrt{t^2-z^2}.
\end{eqnarray}

\section{Zeroth moments of the kinetic equations}
\label{sect:zeromoment}

Integrating Eqs.~(\ref{kineq0}) and (\ref{kineg0}) over three-momentum and including the internal degrees of freedom we obtain the three scalar equations
\begin{eqnarray}
\partial_{\mu } (n_{q}^{\pm} U^{\mu}) &=& 
 \frac{n_{q, \rm eq}^{\pm} - n_{q}^{\pm}}{\tau_{\rm eq}}, 
\label{denq} \\
 \partial_{\mu } (n_{g} U^{\mu} ) &=& 
 \frac{n_{g, \rm eq} - n_{g}}{\tau_{\rm eq}},
\label{deng}
\end{eqnarray}
where we have introduced the non-equilibrium and equilibrium particle densities defined by the expressions
\begin{eqnarray}
n_{q}^{\pm} &=& \frac{g_q}{\pi^2} \frac{e^{\pm \lambda_q/\Lambda_q} \Lambda_q^3}{\sqrt{1+\xi_q}}, 
\quad n_{q, \rm eq}^{\pm} = \frac{g_q}{\pi^2} e^{\pm \mu/T} T^3, 
\label{nq} \\
n_{g} &=& \frac{g_g}{\pi^2} \frac{ \Lambda_g^3}{\sqrt{1+\xi_g}}, 
\quad n_{g, \rm eq} = \frac{g_g}{\pi^2} T^3.
\label{ng}
\end{eqnarray}
Note that for the (0+1)D system we have $U^\mu \partial_\mu = d/d\tau$ and $ \partial_\mu U^\mu = 1/\tau$.

Instead of using Eqs.~(\ref{denq})--(\ref{deng}) we use the difference of the equations for quarks and antiquarks appearing in (\ref{denq}), 
\begin{eqnarray}
&& \frac{d}{d\tau} \left(n_q^+ - n_q^-\right) + \frac{n_q^+ - n_q^-}{\tau} \label{bnc0} \\
&& \hspace{1cm} =  \frac{n_{q, \rm eq}^{+} - n_{q, \rm eq}^{-} - (n_{q}^{+} - n_{q}^{-})}{\tau_{\rm eq}}, \nonumber 
\end{eqnarray}
and the following linear combination of Eqs.~(\ref{denq})--(\ref{deng})
\begin{eqnarray}
&& \alpha \left(\frac{dn_q}{d\tau}  + \frac{n_q}{\tau} \right) +   (1-\alpha) \left(\frac{dn_g}{d\tau} + \frac{n_g}{\tau} \right)
\label{zm0} \\
&& \hspace{1cm} =  \alpha \,\, \frac{n_{q, \rm eq} -n_{q}}{\tau_{\rm eq}} +  (1-\alpha) \,\, \frac{n_{g, {\rm eq}} -n_g}{\tau_{\rm eq}} .
\nonumber 
\end{eqnarray}
In Eq.~(\ref{zm0}) we have introduced the notation $n_q$ and $n_{q, \rm eq} $ for the sum of the quark and antiquark densities,
\begin{eqnarray}
n_q &=& n_q^+ + n_q^- = \frac{2 g_q}{\pi^2} \frac{\cosh( \lambda_q/\Lambda_q) \Lambda_q^3}{\sqrt{1+\xi_q}},
\nonumber \\
n_{q, \rm eq} &=& n_{q, \rm eq}^{+} + n_{q, \rm eq}^{-} = \frac{2 g_q}{\pi^2} \cosh(\mu/T) \,T^3.
\label{nqneq}
\end{eqnarray}

It is important to note that in contrast to Eq.~(\ref{bnc0}), that leads directly to the fundamental law of baryon number conservation, the use of Eq.~(\ref{zm0}) is not so well motivated.  A~straightforward treatment of Eqs.~(\ref{denq})--(\ref{deng}) suggests that Eq.~(\ref{bnc0}) should be supplemented by the two extra equations, for example, one equation for the sum of the quark and antiquark distributions and the other for the gluon distribution. It turns out, however, that the use of three equations obtained from the zeroth moment leads finally to an overdetermined system of equations. Therefore, we have decided to use the linear combination defined by Eq.~(\ref{zm0}), where $\alpha={\rm const}$ is a parameter taken from the range $0 \leq \alpha \leq 1$. 

One may check a posteriori which value of $\alpha$ is the best by comparing the hydrodynamic results with the kinetic-theory results. In this way we have found that the best agreement is obtained for $\alpha=1$. A similar agreement is also obtained for the case $\alpha=0$. We note that the cases $\alpha=0$ and $\alpha=1$ do not introduce the coupling between the quark and gluon sectors at the level of the zeroth moment. This seems to be a desirable situation, since the kinetic equations (\ref{kineq0})--(\ref{kineg0}) treated exactly include the coupling between the quark and gluon sectors {\it only} through the energy-momentum conservation (that is within the first moment of Eqs.~(\ref{kineq0})--(\ref{kineg0})) with the corresponding Landau matching condition, see Ref.~\cite{Florkowski:2014txa}. We come back to the discussion of this point below Eq.~(\ref{eneq}).

\subsection{Baryon number conservation and the corresponding Landau matching}

Equation (\ref{bnc0}) divided by a factor of three gives the constraint on the baryon number density \mbox{$b = \left(n_q^+ - n_q^-\right)/3$}, namely
\begin{eqnarray}
\frac{db}{d\tau} + \frac{b}{\tau} = \frac{b_{\rm eq}-b}{\tau_{\rm eq}}.
\label{bnc1}
\end{eqnarray}
In order to have the baryon number conserved both the left- and right-hand sides of (\ref{bnc1}) should vanish. This yields
\begin{eqnarray}
b(\tau) =  \frac{b_0 \tau_0}{\tau} = \frac{2 g_q}{3\pi^2} \sinh\left(\frac{\lambda_q}{\Lambda_q} \right) 
\frac{\Lambda_q^3}{\sqrt{1+\xi_q}} 
\label{bt}
\end{eqnarray}
and
\begin{eqnarray}
\sinh\left( \frac{\mu}{T} \right) T^3 
= \sinh\left( \frac{\lambda_q}{\Lambda_q} \right) \frac{\Lambda_q^3}{\sqrt{1+\xi_q}},
\label{sinhmuT}
\end{eqnarray}
 respectively. We note that the solution $b(\tau) = b_0 \tau_0/\tau$ has the scaling form known from the Bjorken model~\cite{Bjorken:1982qr}. The quantity $b_0\equiv b(\tau=\tau_0)$ on the right-hand side of Eq.~(\ref{bt}) is the baryon number density at the initial proper time.

Equations (\ref{bt}) and (\ref{sinhmuT}) can be solved for the chemical potentials $\lambda_q$ and $\mu$~\cite{Florkowski:2012as}. We find
\begin{eqnarray}
\frac{\lambda_q}{\Lambda_q} &=&  \sinh^{-1}\left(D \right) =  \ln [D + \sqrt{1+D^2} ]
\label{lambdaq}
\end{eqnarray}
where
\begin{eqnarray}
D(\tau,\Lambda_q,\xi_q) =  \left(\frac{3 \pi^2 b_0 \tau_0 \sqrt{1+\xi_q}}{2 g_q \tau \Lambda_q^3} \right)
\label{D}
\end{eqnarray}
and
\begin{eqnarray}
\frac{\mu}{T} =  \sinh^{-1} \left( \frac{D}{\kappa_q}\right)  =  
 \ln \left[\frac{D}{\kappa_q} + \sqrt{1+\frac{D^2}{\kappa_q^2}} \, \right]
\label{mu}
\end{eqnarray}
where
\begin{eqnarray}
\kappa_q(T,\Lambda_q,\xi_q) = \frac{T^3  \sqrt{1+\xi_q}}{\Lambda_q^3}.
\label{kappaq}
\end{eqnarray}
One should note here that the ratio $D/\kappa_q  = (3\pi^2 b_0 \tau_0)/(2 g_q \tau T^3)$ does not depend on $\Lambda_q$ and $\xi_q$ anymore.
The condition
\begin{equation}
b_{\rm eq} = b,
\label{}
\end{equation}
resulting in Eq.~(\ref{sinhmuT}), should be treated as the Landau matching condition that guarantees the baryon number conservation. It defines the effective baryon chemical potential $\mu$, see Eqs.~(\ref{sinhmuT}) and (\ref{mu}),  in terms of  the variables $\tau, T,\Lambda_q$ and $\xi_q$. Similarly, the baryon conservation equation allows us to express $\lambda_q$ in terms of $\tau, T,\Lambda_q$ and $\xi_q$. In this way, in the following expressions we may completely eliminate both $\mu$ and $\lambda_q$ (the expressions depend, however, in the explicit way on $\tau_0$ and $b_0$).

\subsection{Sum of the zeroth-order moments}

Using Eqs.~(\ref{lambdaq}) and (\ref{mu}) and the mathematical identity 
\begin{equation}
\cosh[\sinh^{-1}(x)] = \sqrt{1+x^2}
\end{equation}
we find the expressions for the quark non-equilibrium and equilibrium densities
\begin{eqnarray}
n_q &=& \frac{2 g_q}{\pi^2} \frac{\sqrt{1+D^2 } \Lambda_q^3}{\sqrt{1+\xi_q}},
\nonumber \\
n_{q, \rm eq} &=& \frac{2 g_q}{\pi^2}  \sqrt{1+ D^2/\kappa_q^2} \,\,\,T^3,
\label{nqneq}
\end{eqnarray}
respectively. 
Using this notation we rewrite Eq.~(\ref{zm0}) in the form
\begin{eqnarray}
&&\frac{d}{d\tau}  \left(\alpha \frac{ \sqrt{1+D^2}  \Lambda_q^3}{\sqrt{1+\xi_q}} 
+ (1-\alpha)   \frac{{\tilde r} \Lambda_g^3}{\sqrt{1+\xi_g}} \right)
\nonumber \\
&&+  \left( \frac{1}{\tau} + \frac{1}{\tau_{\rm eq}} \right) 
\left( \alpha \frac{\sqrt{1+D^2}  \Lambda_q^3}{\sqrt{1+\xi_q}} + (1-\alpha)  \frac{ {\tilde r}\Lambda_g^3}{\sqrt{1+\xi_g}}  \right)
\nonumber \\
&&  = \frac{T^3}{\tau_{\rm eq}}  \left( \alpha \sqrt{1+ D^2/\kappa_q^2}+ (1-\alpha) {\tilde r} \right),
\label{3zero}
\end{eqnarray}
where ${\tilde r} $ is the ratio of the internal degrees of freedom for gluons and quarks 
\begin{eqnarray}
\tilde{r}=\frac{g_g}{2g_q}.
\end{eqnarray}
In the numerical calculations we use the value $\tilde{r}=2/3$.

\section{Energy-momentum conservation law}
\label{sect:firstmoment}

\subsection{Landau matching for the energy density}

The first moment of the sum of the kinetic equations (\ref{kineq0}) and (\ref{kineg0}) gives the divergence of the energy-momentum tensor
\begin{eqnarray}
\partial_\mu T^{\mu \nu} =  \frac{1}{\tau_{\rm eq}} \,
U_{\mu} \left(T_{q, \rm eq}^{\mu \nu}+T_{g, \rm eq}^{\mu \nu}-(T_q^{\mu \nu}+T_g^{\mu \nu})\right),
\label{enmomcon0}
\end{eqnarray}
where the energy-momentum tensor $T^{\mu \nu}$ includes the quark and gluon contributions
\begin{eqnarray}
T^{\mu \nu}=T_q^{\mu \nu}+T_g^{\mu \nu}.
\label{Tmunuqgp}
\end{eqnarray}
The Lorentz structure of the distribution functions for quarks and gluons implies the following forms of the energy-momentum tensors for quarks and gluons~\cite{Florkowski:2010cf}
\begin{eqnarray}
T_i^{\mu \nu} &=& (\varepsilon_i+P_{i,T}) U^\mu U^\nu - P_{i, T} g^{\mu \nu} \nonumber \\
&& -(P_{i,T}-P_{i,L}) Z^\mu Z^\nu \, ,
\label{Tmunu}
\end{eqnarray}
 where $\varepsilon_i$, $P_{i,L}$ and $P_{i,T}$ denote energy density, longitudinal and transverse pressure, respectively, and the index $i$ stands for quarks ($i=q$) or gluons ($i=g$). Here the energy densities of quarks and gluons are given by the expressions
\begin{eqnarray}
\varepsilon_q &=& \frac{6 g_q \Lambda^4_q}{\pi^2}  \sqrt{1+D^2} \,\,{\cal R}(\xi_q),
\nonumber \\
\varepsilon_g &=& \frac{3 g_g \Lambda^4_g}{\pi^2} \,\,{\cal R}(\xi_g),
\label{epsilon} 
\end{eqnarray}
where the function ${\cal R}(\xi)$ is defined as~\cite{Martinez:2010sc}
\begin{equation}
{\cal R}(\xi) = \frac{1}{2(1+\xi)} 
\left[1+ \frac{ (1+\xi) \tan^{-1} \sqrt{\xi}} {\sqrt{\xi} } \right].
\end{equation}
Correspondingly, for the equilibrium part we find
\begin{eqnarray}
T^{\mu \nu}_{\rm eq}=T_{q, \rm eq}^{\mu \nu}+T_{g, \rm eq}^{\mu \nu},
\label{Tmunuqgp}
\end{eqnarray}
where
\begin{equation}
T_{i, \rm eq}^{\mu \nu} = (\varepsilon_{i, \rm eq}+P_{i, \rm eq}) U^\mu U^\nu - P_{i, \rm eq} g^{\mu \nu},
\label{Tmunueq}
\end{equation}
with $\varepsilon_{i, \rm eq}$ and $P_{i, \rm eq}$ being equilibrium energy density and pressure, respectively. The equilibrium energy densities for quarks and gluons are given by the expressions
\begin{eqnarray}
\varepsilon_{q, \rm eq} &=& \frac{6 g_q T^4}{\pi^2} \sqrt{1+ D^2/\kappa_q^2}, \nonumber \\
\varepsilon_{g, \rm eq} &=& \frac{3 g_g T^4}{\pi^2}. 
\label{epsiloneq} 
\end{eqnarray}

The energy-momentum conservation law $\partial_\mu T^{\mu\nu} = 0$ requires that the right-hand side of Eq.~(\ref{enmomcon0}) vanishes.  This is nothing else but the Landau matching condition for the energy-momentum conservation. This matching requires that the energy determined from  the non-equilibrium distribution functions is the same as the energy obtained with the equilibrium distribution functions
\begin{eqnarray}
\varepsilon = \varepsilon_q + \varepsilon_g = \varepsilon_{\rm eq} = \varepsilon_{q, \rm eq} + \varepsilon_{g, \rm eq}.
\label{LM0}
\end{eqnarray}
This leads directly to the constraint on the effective temperature $T$,
\begin{eqnarray}
T^4=  \, \frac{ \Lambda^4_q \sqrt{1+D^2} \, {\cal R}(\xi_q)  + \Lambda^4_g{\tilde r} 
\, {\cal R}(\xi_g) }{ \sqrt{1+D^2/\kappa_q^2} + {\tilde r}}.
\label{TL}
\end{eqnarray}

\subsection{Energy and momentum conservation}

In the (0+1)D case considered here the energy and momentum conservation takes the form 
\begin{equation}
\frac{d\varepsilon}{d\tau} = -\frac{\varepsilon+P_L}{\tau},
\label{enmomcon01}
\end{equation}
where $P_L$ is the sum of the longitudinal pressures for quarks and gluons,
\begin{eqnarray}
P_{q,L} &=& \frac{6 g_q \Lambda^4_q}{\pi^2}  \sqrt{1+D^2} \,\,{\cal R}_L(\xi_q),
\nonumber \\
P_{g,L} &=& \frac{3 g_g \Lambda^4_g}{\pi^2} \,\,{\cal R}_L(\xi_g),
\label{epsilon} 
\end{eqnarray}
with ${\cal R}_L$ defined through the formula
\begin{equation}
{\cal R}_L(\xi) =- \left[2(1+\xi)R^\prime(\xi)+R(\xi) \right].
\end{equation}
This leads directly to the formula
\begin{eqnarray}
&&\frac{d}{d\tau} \left[    \Lambda_q^4 \sqrt{1+D^2} \,
{\cal R}(\xi_q)  + {\tilde r} \Lambda_g^4 {\cal R}(\xi_g) \right] \label{eneq} \\
&& = \frac{2}{\tau} \left[ \Lambda^4_q \sqrt{1+D^2} \, (1+\xi_q) 
{\cal R}^\prime (\xi_q)  + {\tilde r}  \Lambda^4_g (1+\xi_g)  {\cal R}^\prime (\xi_g) \vphantom{e^{\lambda/\Lambda}} \right].
 \nonumber
\end{eqnarray}

We note that Eqs.~(\ref{TL}) and (\ref{eneq}) couple the quark and gluon parameters in the similar way as these two sectors are coupled in the exact treatment of the kinetic equations~(\ref{kineq0})--(\ref{kineg0}). Hence, the best agreement between the hydrodynamic equations and the kinetic theory {\it may} be expected if no other coupling is incorporated into the hydrodynamic approach. This, in turn, suggests to use the values $\alpha=1$~or~$\alpha=0$ in Eq.~(\ref{zm0}). Such a conjecture has been supported by our numerical calculations done with various values of $\alpha$ taken within the range $0 \leq \alpha \leq 1$.

\section{Second moment of the kinetic equation}
\label{sect:secondmoment}

So far we have constructed three equations, see Eqs.~(\ref{3zero}), (\ref{TL}), and (\ref{eneq}), for five unknown functions: $\Lambda_q$, $\Lambda_g$, $\xi_q$, $\xi_g$, and $T$. In order to close the system of equations we need to include two extra equations. We shall construct them using the second moment of the kinetic equation.

The second moment of the kinetic equation in the relaxation time approximation was studied in Ref.~\cite{Tinti:2013vba}, where a boost-invariant and cylindrically symmetric systems were analyzed. In this case, it was shown that it is convenient to select the following equations as the basis for the hydrodynamic approximation 
\begin{eqnarray}
&& \frac{d}{d\tau} \ln\Theta_I + \theta - 2\theta_I -\frac{1}{3}\sum_J \left[\frac{d}{d\tau} \ln\Theta_J + \theta - 2\theta_J \right]  \nonumber \\
&& = \frac{1}{\tau_{\rm eq}} \left[ \frac{\Theta_{\rm eq}}{\Theta_I} - 1 \right] -\frac{1}{3}\sum_J \left\{ \frac{1}{\tau_{\rm eq}} \left[ \frac{\Theta_{\rm eq}}{\Theta_J} - 1 \right]  \right\} \,.
\label{sum}
\end{eqnarray} 
Here $I=X,Y,Z$ and $J=X,Y,Z$ are space indices. In the (0+1)D case the coefficients $\theta_I$ have the form: $\theta_X=\theta_Y=0$, \mbox{$\theta_Z=-1/\tau$}, and $\theta=1/\tau$. The three functions $\Theta_I$ are obtained as  contractions of the third moment of the distribution function with the tensor $U \otimes I \otimes I$ (where $I$ is now the four-vector corresponding to the index $I$). The function $\Theta_{\rm eq}$ is obtained by the contraction of the third moment of the equilibrium distribution function with any of  the tensors $U \otimes I \otimes I$ (the result is independent of $I$). The four-vectors $U$ and $Z$ are defined above. The four vectors  $X$ and $Y$ in our case, where the transverse expansion is neglected, are given simply by the formulas~\cite{Florkowski:2011jg}
\begin{equation}
X^\mu = (0,1,0,0), \quad Y^\mu=(0,0,1,0).
\label{XY}
\end{equation}
It is important to emphasise that in our one-dimensional case only one out of three equations in (\ref{sum}) is independent~\footnote{By construction, only two equations in (\ref{sum}) are independent. In the case where the transverse flow is neglected, these two become degenerate.}. It may be taken as
\begin{eqnarray}
\frac{d}{d\tau}\ln\Theta_X-\frac{d}{d\tau}\ln\Theta_Z-\frac{2}{\tau}=\frac{\Theta_{\rm eq}}{\tau_{\rm eq}}\left[\frac{1}{\Theta_X}-\frac{1}{\Theta_Z}  \right].
\label{sumX}
\end{eqnarray}
Since Eqs.~(\ref{sum}) turned out to be very successful in the construction of hydrodynamic models \cite{Nopoush:2015yga}, in particular, they are consistent with the Israel-Stewart theory for systems close to equilibrium, we use this form separately for quarks and gluons. See also our remarks below Eq.~(\ref{eneq}).

\subsection{Quarks and antiquarks}

Following the method of Ref.~\cite{Florkowski:2014bba} one can derive the following formulas for the sum of the quark and antiquark distributions
\begin{eqnarray}
\Theta_X^{q} &=&\Theta_Y^{q}=\frac{8g_q\Lambda^5_q}{\pi^2(1+\xi_q)^{1/2}}\sqrt{1+D^2},
\nonumber \\
\Theta_Z^{q} &=& \frac{8g_q\Lambda^5_q}{\pi^2(1+\xi_q)^{3/2}}\sqrt{1+D^2}.
\label{ThetaqXYZ}
\end{eqnarray}
Similarly, for the equilibrium quark functions one gets
\begin{eqnarray}
\Theta_{X,\rm eq}^{q}=\Theta_{Y,\rm eq}^{q}=\Theta_{Z,\rm eq}^{q}
=\frac{8g_q T^5}{\pi^2}  \sqrt{1+D^2/\kappa_q^2}.
\nonumber \\
\label{ThetaqXYZeq}
\end{eqnarray}
Using Eqs.~(\ref{ThetaqXYZ}) and (\ref{ThetaqXYZeq}) in (\ref{sumX}) we find
\begin{eqnarray}
&&\frac{d}{d\tau}\ln\left(\frac{\Lambda_q^5}{(1+\xi_q)^{1/2}}\sqrt{1+D^2}\right) 
\label{Tintiq} \\
&-& \frac{d}{d\tau}\ln\left(\frac{\Lambda_q^5}{(1+\xi_q)^{3/2}}\sqrt{1+D^2}\right)-\frac{2}{\tau} 
\nonumber \\
&=& \frac{T^5}{\tau_{\rm eq}\Lambda^5_q}\xi_q(1+\xi_q)^{1/2}\frac{\sqrt{1+D^2/\kappa_q^2}}{\sqrt{1+D^2}}.
\nonumber 
\end{eqnarray}

\begin{figure}[t!]
\includegraphics[angle=0,width=0.4125\textwidth]{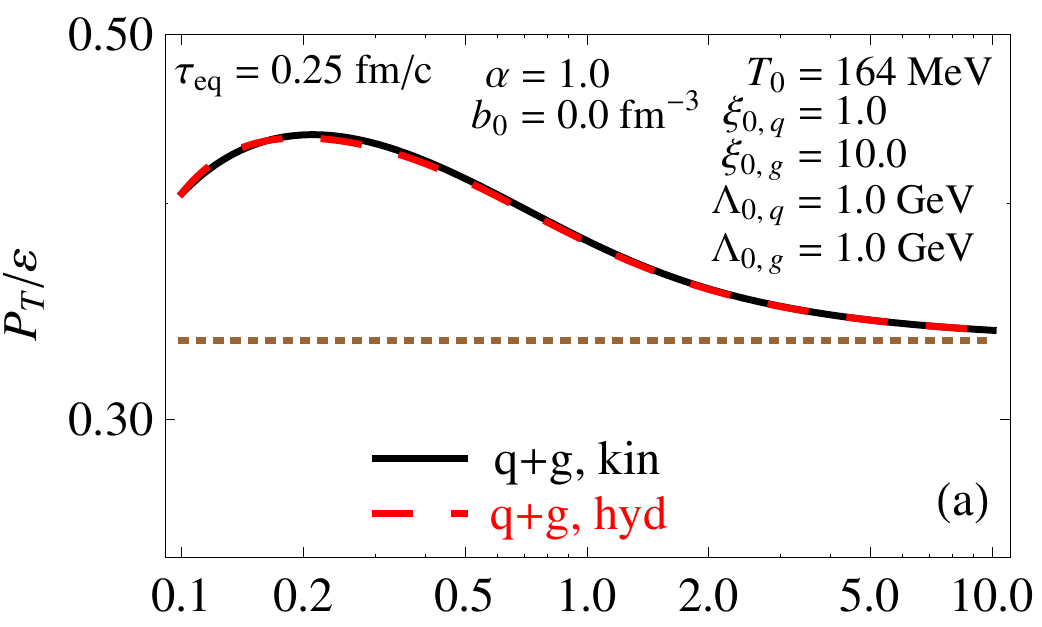} \\
\includegraphics[angle=0,width=0.4125\textwidth]{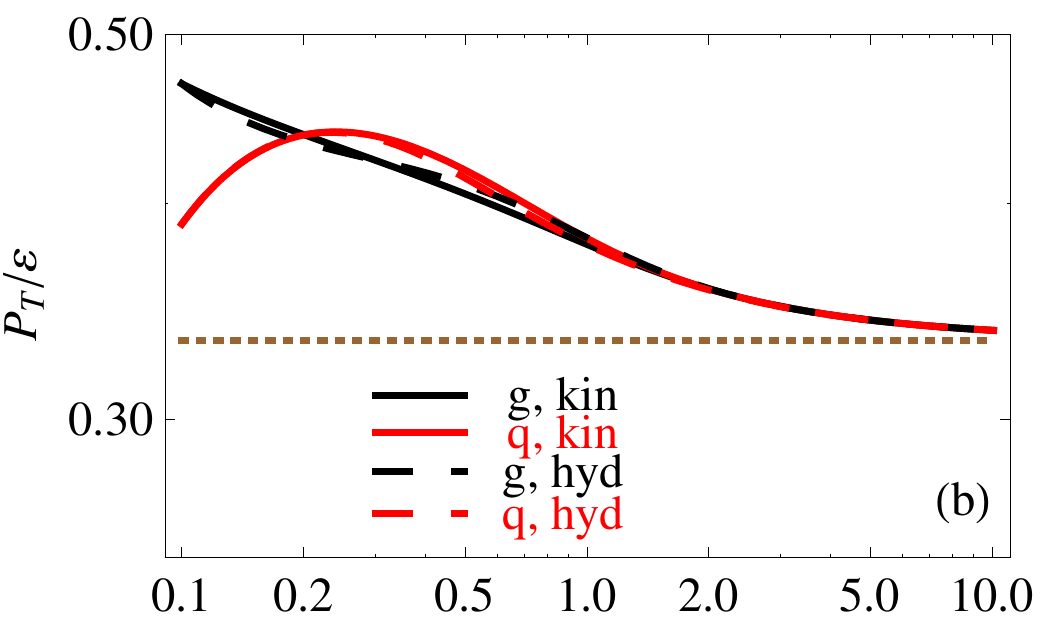}  \\
\includegraphics[angle=0,width=0.4125\textwidth]{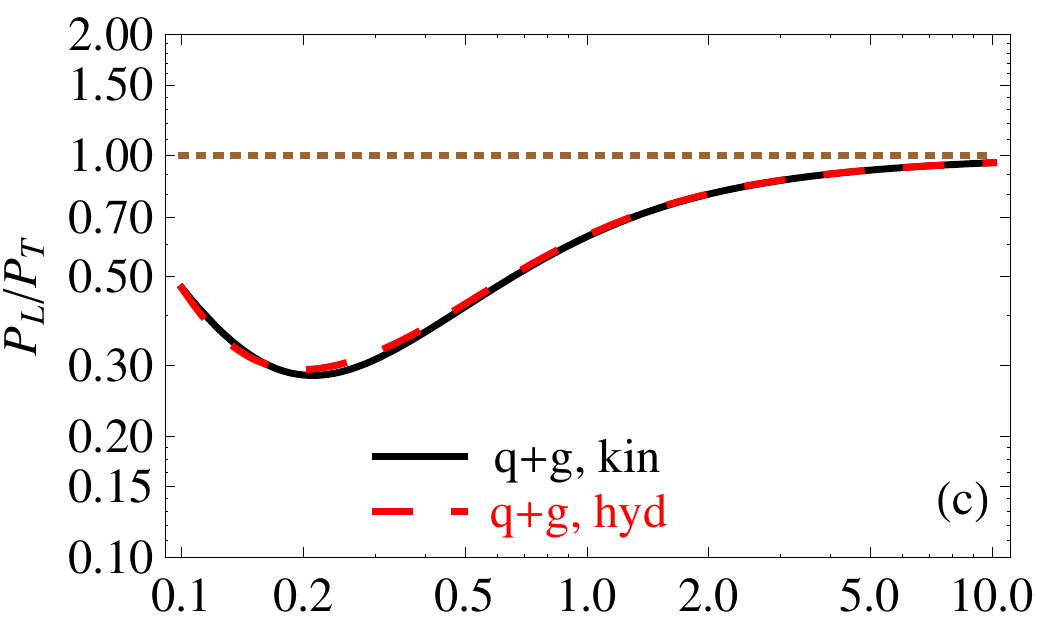} \\
\includegraphics[angle=0,width=0.4125\textwidth]{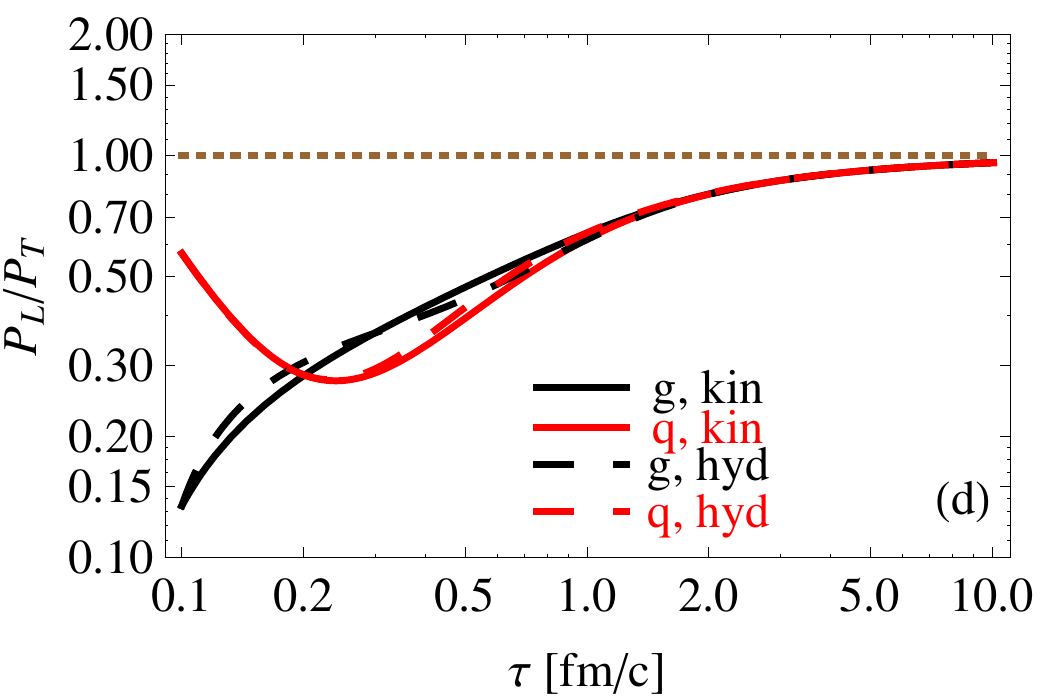}  \\
\caption{(Color online) Comparison of the hydrodynamic  and kinetic-theory results for the initial oblate-oblate configurations. Detailed description in the text. }
\label{fig:oo}
\end{figure}

\subsection{Gluons}

In the case of gluons, the analogous expressions are
\begin{eqnarray}
\Theta_X^g &=& \Theta_Y^g=\frac{4g_g\Lambda^5_g}{\pi^2(1+\xi_g)^{1/2}}  ,
\nonumber \\
\Theta_Z^g &=& \frac{4g_g\Lambda^5_g}{\pi^2(1+\xi_g)^{3/2}},
\label{ThetagXYZ}
\end{eqnarray}
and the equilibrium functions are
\begin{eqnarray}
\Theta_{X,\rm eq}^g=\Theta_{Y,\rm eq}^g=\Theta_{Z,\rm eq}^g=\frac{4g_g T^5}{\pi^2}.
\label{ThetagXYZeq}
\end{eqnarray}

Using Eqs.~(\ref{ThetagXYZ}) and (\ref{ThetagXYZeq}) in (\ref{sumX}) one obtains
\begin{eqnarray}
&&\frac{d}{d\tau}\ln\left(\frac{\Lambda_g^5}{(1+\xi_g)^{1/2}}\right) 
- \frac{d}{d\tau}\ln\left(\frac{\Lambda_g^5}{(1+\xi_g)^{3/2}}\right)-\frac{2}{\tau} \
\nonumber \\
&&  = \frac{T^5}{\tau_{\rm eq}\Lambda^5_g}\xi_g(1+\xi_g)^{1/2}. 
\label{Tintig}
\end{eqnarray}
Equations~(\ref{3zero}), (\ref{TL}), and (\ref{eneq}) together with Eqs.~(\ref{Tintiq}) and (\ref{Tintig}) represent five independent equations that allow us to determine five unknown functions of the proper time: $\Lambda_q$, $\Lambda_g$, $\xi_q$, $\xi_g$, and $T$. 

\begin{figure}[t!]
\includegraphics[angle=0,width=0.4125\textwidth]{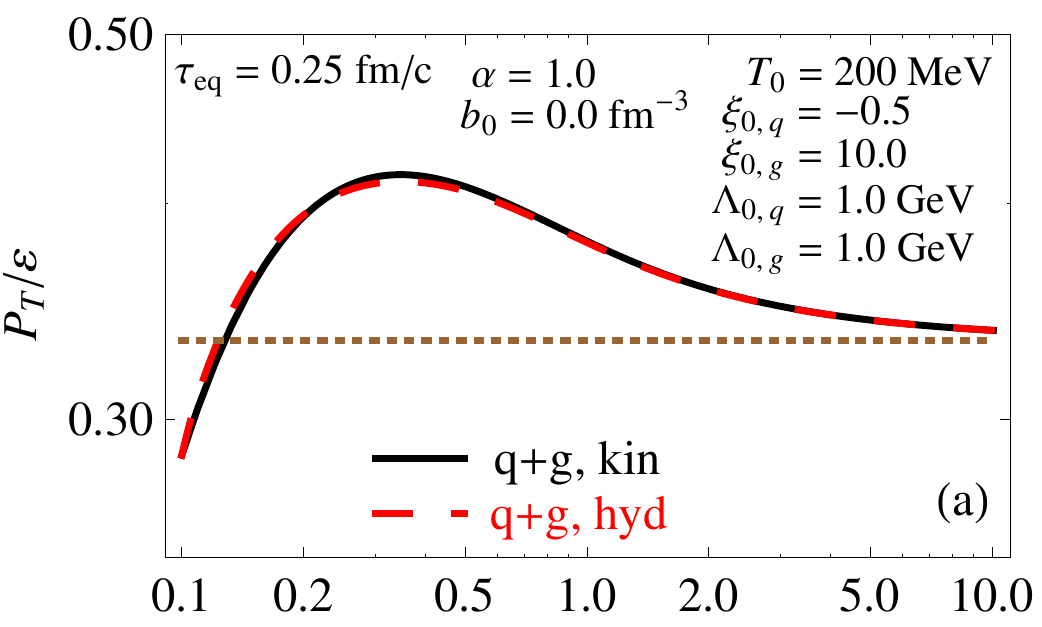} \\
\includegraphics[angle=0,width=0.4125\textwidth]{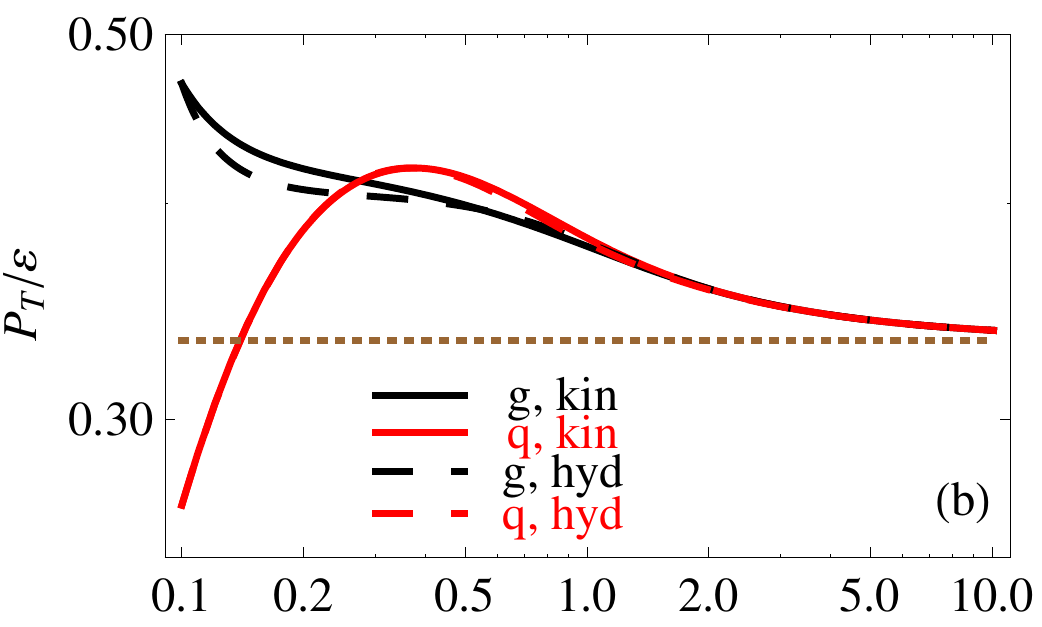}  \\
\includegraphics[angle=0,width=0.4125\textwidth]{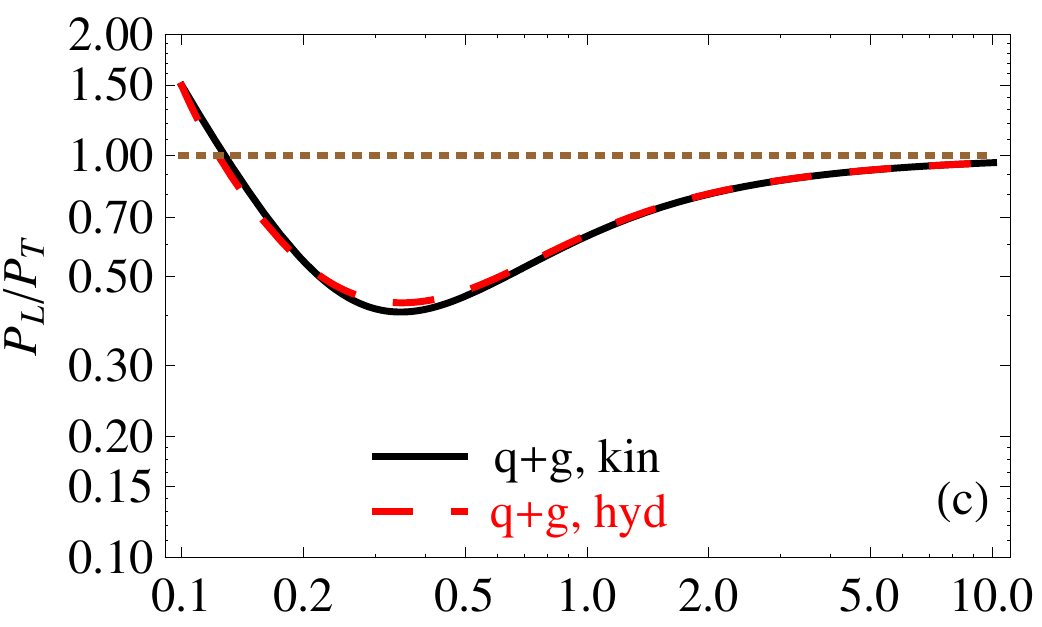} \\
\includegraphics[angle=0,width=0.4125\textwidth]{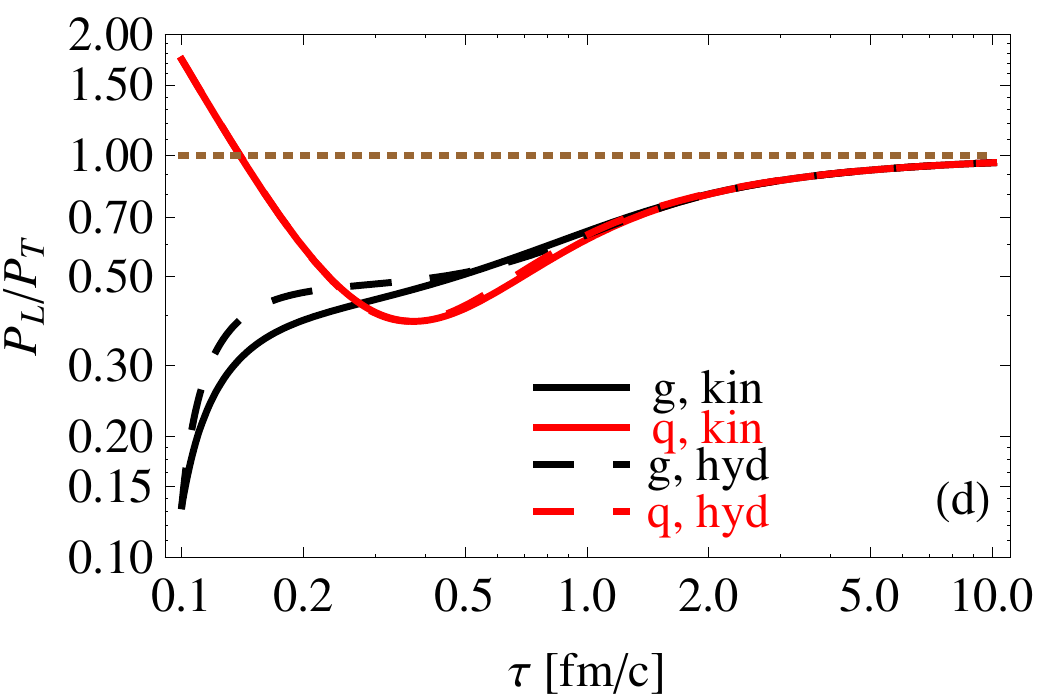}  \\
\caption{(Color online) The same as Fig.~\ref{fig:oo} but for the prolate-oblate configuration. }
\label{fig:po}
\end{figure}

\section{Results}
\label{sect:res}

In this Section we present our numerical results for four different types of initial conditions. In the first case the initial conditions correspond to the oblate quark and gluon distribution functions (oblate-oblate configuration). In the second case the initial distribution of quarks is prolate, while the gluon distribution is oblate (prolate-oblate configuration), and in the third case the two distributions are prolate (prolate-prolate configuration). In the first three cases we set baryon number density equal to zero. The effect of the non-vanishing baryon density is studied in the fourth case where the quark and gluon distributions are both oblate.

We note that the oblate (prolate) distributions correspond to positive (negative) values of the anisotropy parameter $\xi$ and, consequently, to the transverse pressure larger (smaller) than the longitudinal pressure. The results of the microscopic calculations indicate that the initial conditions in relativistic heavy-ion collisions correspond to the (oblate) case where $P_T > P_L$,  as soon as the coherent longitudinal colour fields disappear~\cite{Ryblewski:2013eja,Gelis:2013rba,Ruggieri:2015yea}.  This suggests that the oblate-oblate configuration is probably the most realistic one.

In all the considered cases the initial starting (proper) time is $\tau=\tau_0=0.1$~fm/c and we continue the evolution till $\tau=$~10~fm/c. The initial transverse-momentum parameters $\Lambda_{0,i}$ for quarks and gluons have been set equal to 1~GeV. From the Landau matching condition for the energy, see Eq.~(\ref{TL}), we determine the initial temperature $T_0$ which is different for the cases with different initial anisotropies. The value of the relaxation time used in this work is constant, $\tau_{\rm eq} = 0.25$~fm/c. A temperature dependent $\tau_{\rm eq}$ can be also used, as it is described in~\cite{Florkowski:2013lza,Florkowski:2013lya}. However, to check the agreement with the kinetic theory it is enough to use a constant value.

In the four cases presented here, the results of anisotropic hydrodynamics are compared with the exact solutions of the kinetic equations (\ref{kineq0})--(\ref{kineg0}) that have been constructed for (0+1)D systems in Ref.~\cite{Florkowski:2014txa}. We refer to this paper for all details connected with the exact treatment of Eqs.~(\ref{kineq0})--(\ref{kineg0}). Here we only emphasise that the initial distribution functions used in the kinetic-theory calculations are specified by the same initial parameters as those used in the hydrodynamic calculations. 

\begin{figure}[t!]
\includegraphics[angle=0,width=0.4125\textwidth]{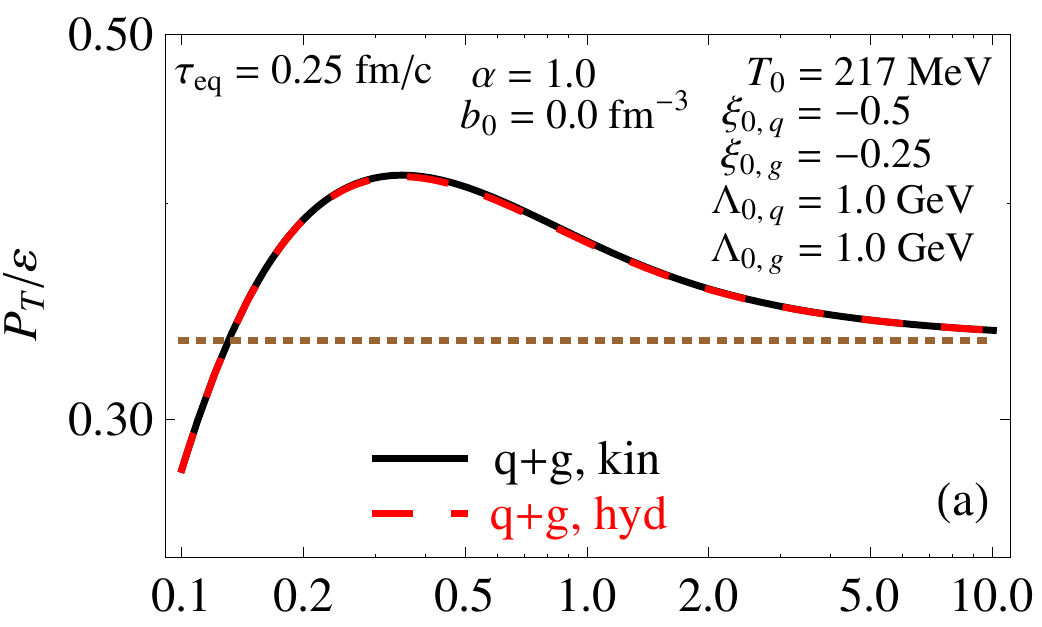} \\
\includegraphics[angle=0,width=0.4125\textwidth]{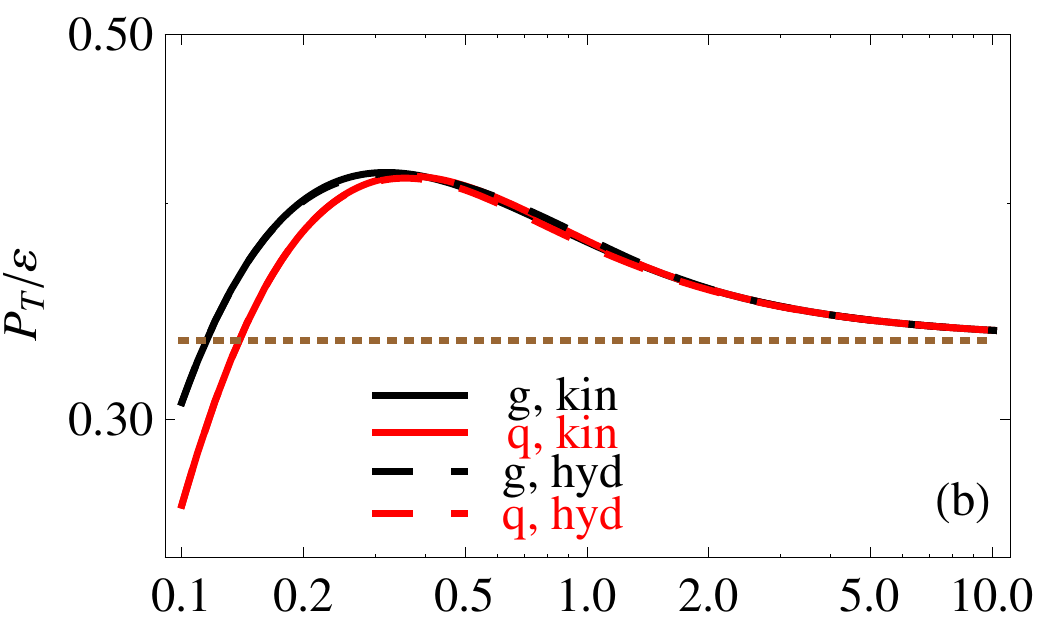}  \\
\includegraphics[angle=0,width=0.4125\textwidth]{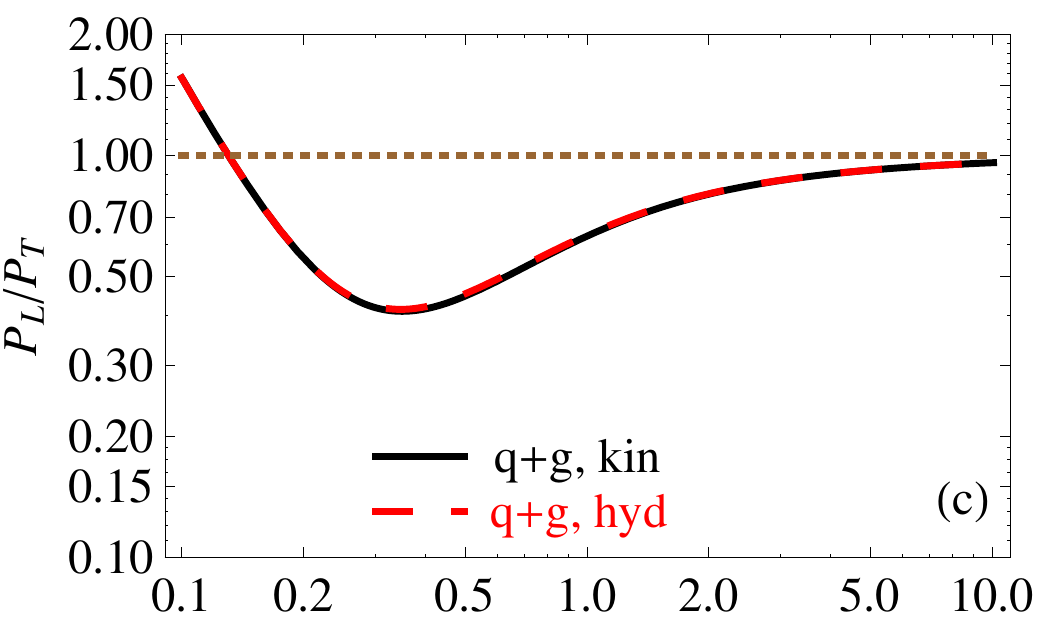} \\
\includegraphics[angle=0,width=0.4125\textwidth]{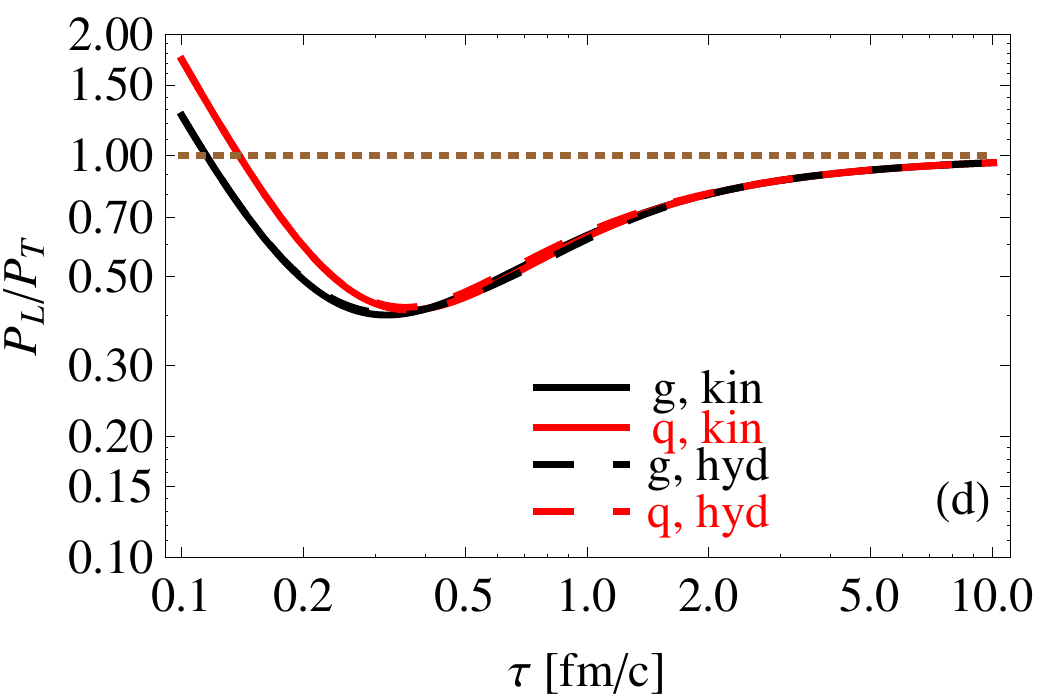}  \\
\caption{(Color online) The same as Fig.~\ref{fig:oo} but for the prolate-prolate configuration.}
\label{fig:pp}
\end{figure}

The results for the oblate-oblate initial configuration with $\xi_{0,q}=1$ and $\xi_{0,g}=10$ are shown in Fig.~\ref{fig:oo}. The panel (a) shows the ratio of the total transverse pressure to the total energy density, $P_T/\varepsilon$. The solid line describes the result of the kinetic theory, while the dashed  line is the result of anistropic hydrodynamics. The panel (b) shows the same ratio but for the individual quark (red) and gluon (black) components. The panels (c)  and (d) show the  $P_L/P_T$ ratio for the whole system and individual components, respectively. Figures \ref{fig:po} and \ref{fig:pp} show the same ratios as those presented in Fig.~\ref{fig:oo} but for the prolate-oblate ($\xi_{0,q} = -0.5$, $\xi_{0,g}=10$) and prolate-prolate  ($\xi_{0,q} = -0.5$, $\xi_{0,g}=-0.25$)  initial conditions, respectively. The coding of the lines is the same as in Fig.~\ref{fig:oo}. 

\begin{figure}[t!]
\includegraphics[angle=0,width=0.4125\textwidth]{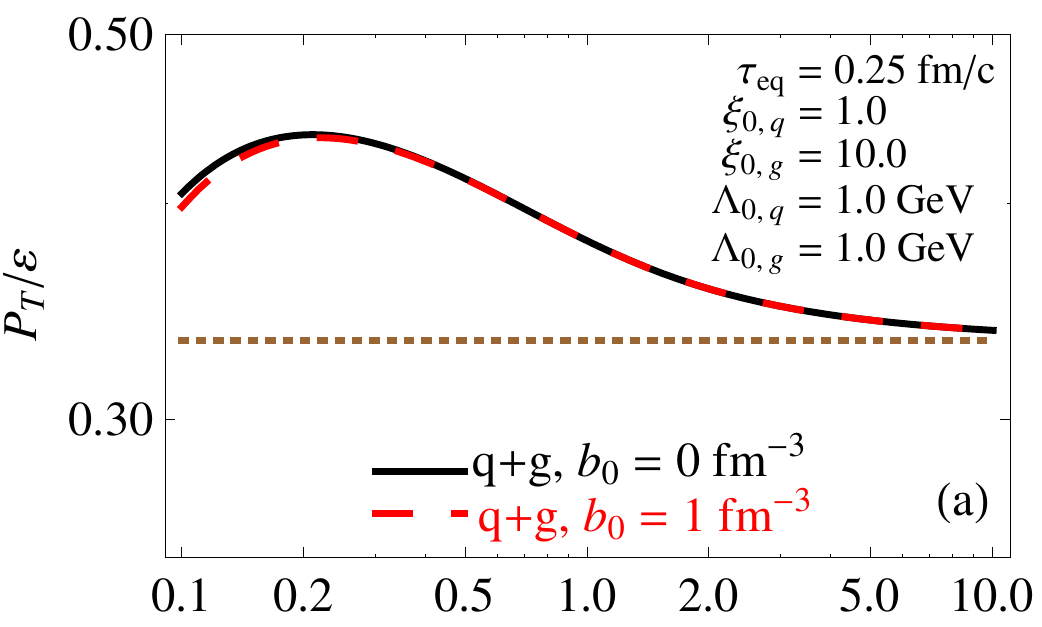} \\
\includegraphics[angle=0,width=0.4125\textwidth]{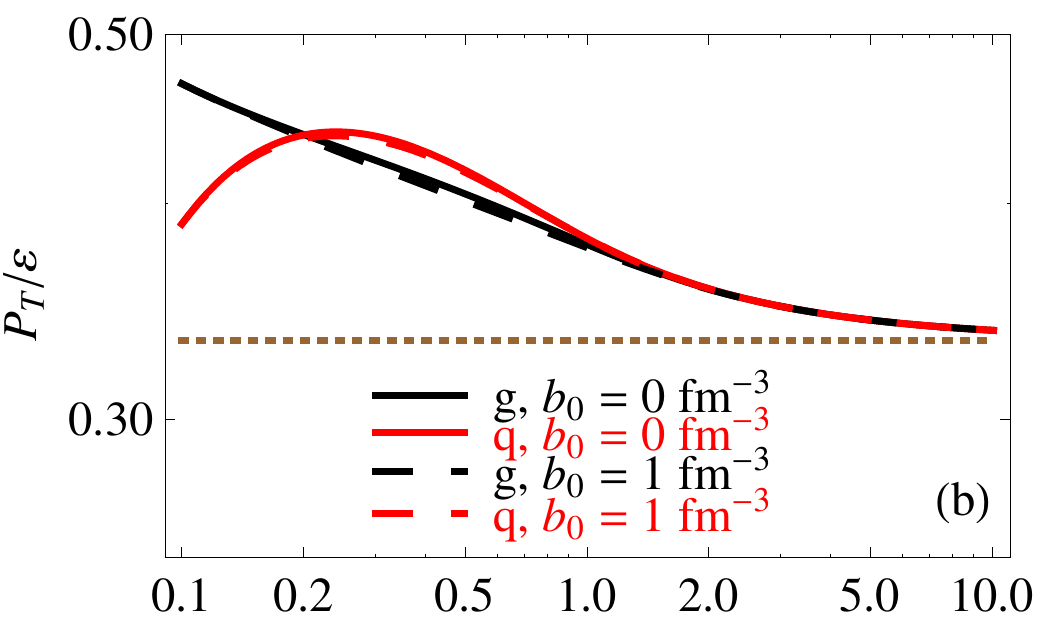}  \\
\includegraphics[angle=0,width=0.4125\textwidth]{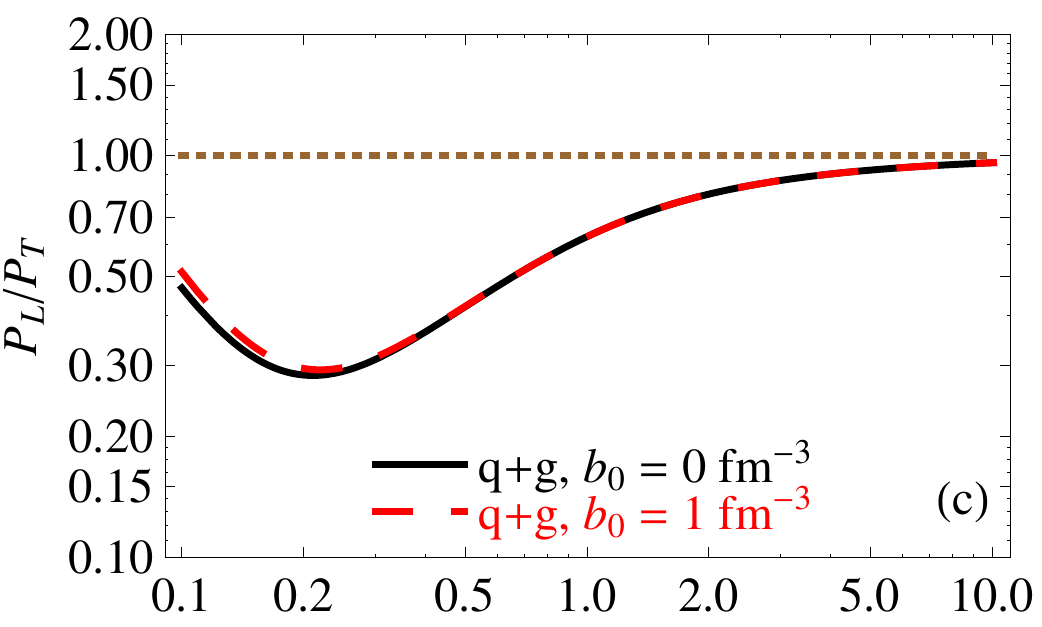} \\
\includegraphics[angle=0,width=0.4125\textwidth]{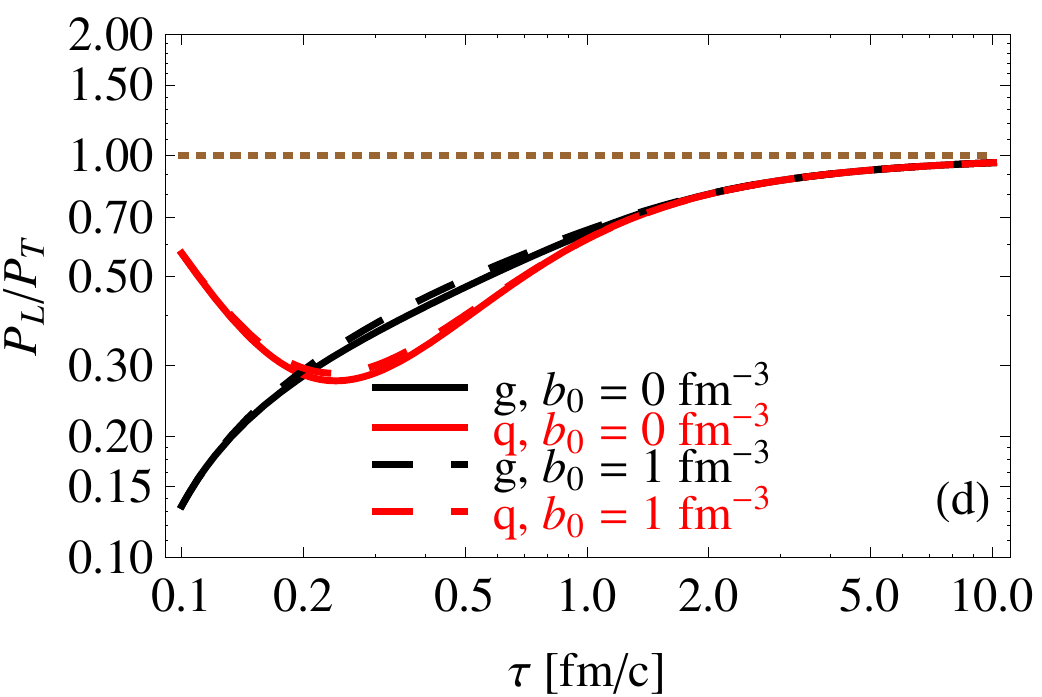}  \\
\caption{(Color online) Comparison of the results obtained with zero and finite initial baryon density for the oblate-oblate configuration.}
\label{fig:oob}
\end{figure}

The results presented in Fig.~\ref{fig:oo} show that the hydrodynamic description agrees very well with the kinetic results. One may notice, however, that the best agreement is achieved for total quantities that include both quarks and gluons. On the other hand, the worst agreement is obtained for gluons alone. This situation changes if we set $\alpha=0$ instead of $\alpha=1$, however,  the overall results for the case $\alpha=1$ are the best. In Figs.~\ref{fig:po}~and~\ref{fig:pp} again a good agreement between the hydrodynamic and kinetic-theory results can be seen, especially for the prolate-prolate configurations. For the prolate-oblate case, the agreement is a bit worse, but we have to keep in mind that such an initial configuration is extremely out of equilibrium --- not only the two distributions are highly anisotropic in the momentum space, but their individual types of anisotropy are different. A correct description of such a non-equilibrium case within a hydrodynamic approach is challenging.

Finally, in Fig.~\ref{fig:oob} we show the effect of finite baryon density. The calculations are done for the two cases: $b_0 = 0$ and $b_0 = 1~\hbox{fm}^{-3}$. In these two cases the initial oblate-oblate configuration is assumed. In agreement with earlier studies we find that the effect of finite baryon density is very small, unless the initial baryon number density is extremely large (\mbox{$b_0 \gg 1~\hbox{fm}^{-3}$}).

We emphasise that  the results shown in Figs.~\ref{fig:oo}--\ref{fig:oob} have been obtained with $\alpha=1$ in Eq.~(\ref{zm0}), which has turned out to be the best choice. Similar agreement has been also obtained for the case $\alpha=0$. On the other hand, the case $\alpha=0.5$  gives much worse agreement.

\bigskip
\section{Conclusions}
\label{sect:con}

We have constructed a new set of equations for anisotropic hydrodynamics describing a mixture of anisotropic quark and gluon fluids. The consistent treatment of the zeroth, first and second moments of the kinetic equations allows us to construct our approach with more general forms of the anisotropic phase-space distribution functions than those used in similar earlier studies~ \cite{Florkowski:2012as,Florkowski:2013uqa,Florkowski:2014txa}. In this way, the main problems of the previous formulations of anisotropic hydrodynamics for mixtures have been overcome and the good agreement with the exact kinetic-theory results is obtained.

\bigskip
\begin{acknowledgments}

We thank Michael Strickland for stimulating and clarifying discussions.
Research supported in part by Polish National Science Center grants No. DEC-2012/05/B/ST2/02528 (WF), No. DEC-2012/06/A/ST2/00390 (WF, EM, LT), and DEC-2012/07/D/ST2/02125 (RR). 

\end{acknowledgments}

\bigskip

\end{document}